\documentclass[prx,twocolumn,superscriptaddress]{revtex4}
\usepackage{amsmath,amssymb,mathrsfs}

\usepackage{hyperref}
\usepackage{paralist}
\usepackage{pdfpages}
 \usepackage{graphicx}

\def\be{\begin{equation}}
\def\ee{\end{equation}}
\def\bea{\begin{eqnarray}}
\def\eea{\end{eqnarray}}

\begin{document}

\title{Dimensionless ratios: characteristics of quantum liquids and their phase transitions }

\author{Yi-Cong Yu}
\affiliation{State Key Laboratory of Magnetic Resonance and Atomic and Molecular Physics,
Wuhan Institute of Physics and Mathematics, Chinese Academy of Sciences, Wuhan 430071, China}

\author{Yang-Yang Chen}
\affiliation{State Key Laboratory of Magnetic Resonance and Atomic and Molecular Physics,
Wuhan Institute of Physics and Mathematics, Chinese Academy of Sciences, Wuhan 430071, China}


\author{Hai-Qing Lin}
\email[e-mail:]{haiqing0@csrc.ac.cn}
\affiliation{Beijing Computational Science Research Center, Beijing 100094, China}

\author{Rudolf A.\ R\"{o}mer}
\affiliation{Department of Physics and Centre for Scientific Computing, University of Warwick,a Coventry, CV4 7AL, UK}

\author{Xi-Wen Guan}
\email[e-mail:]{xiwen.guan@anu.edu.au}
\affiliation{State Key Laboratory of Magnetic Resonance and Atomic and Molecular Physics,
Wuhan Institute of Physics and Mathematics, Chinese Academy of Sciences, Wuhan 430071, China}
\affiliation{Center for Cold Atom Physics, Chinese Academy of Sciences, Wuhan 430071, China}
\affiliation{Department of Theoretical Physics, Research School of Physics and Engineering,
Australian National University, Canberra ACT 0200, Australia}

\date{$Revision: 1.2 $, compiled \today}

\begin{abstract}
Dimensionless ratios of physical properties can characterize low-temperature phases in a wide variety of materials. As such, the Wilson ratio (WR), the Kadowaki-Woods ratio and the Wiedemann\--Franz law capture essential features of Fermi liquids in metals, heavy fermions, etc.
Here we prove that the phases of many-body interacting multi-component quantum
liquids in one dimension (1D)  can be described by WRs based on the
compressibility, susceptibility and specific heat associated with each
component.
These WRs arise due to additivity rules within subsystems reminiscent of the rules for multi-resistor
networks in series and parallel --- a novel and useful characteristic of multi-component Tomonaga-Luttinger liquids (TLL)   independent of microscopic details of the systems. 
Using  experimentally realized
multi-species cold atomic gases as examples, we prove that the Wilson ratios
uniquely identify phases of TLL, while providing universal
scaling relations at the boundaries between phases.
Their values within a phase   are solely determined by the stiffnesses and sound velocities of subsystems and identify the 
internal degrees of freedom of said phase such as its spin-degeneracy.
This finding can be directly applied to a wide range of 1D  many-body systems and reveals deep physical insights into recent experimental measurements of the universal thermodynamics in ultracold atoms and spins. 
\end{abstract}
\maketitle

\section{Introduction}

One of the central challenges in condensed matter physics is to understand how
different phases of matter can arise and how these phases can be characterized.
Many phenomena such as, e.g., superconductivity, magnetism and quantum phase transitions
 in strongly correlated systems \cite{Sac99}, Bose-Einstein condensation of dilute
gases \cite{PetS08} and of excitons in semiconductors, electronic transport in low-dimensional
systems and heavy-fermion physics \cite{Hew97}, are known to exist due to the
\emph{collective} nature of the underlying many-body processes. 
Collective
phenomena are particularly strong in low-dimensional systems where the reduced
dimensionality enhances the interaction of elementary constituents \cite{Gia04,Ess05,CazCGO11,ImaSG12}.
In order to characterize the various phases, dimensionless  ratios such as the
celebrated Wiedemann-Franz (WF) law \cite{WieF53} or the Kadowaki-Wood
ratio \cite{KadW86,JacFP09} are very useful. They usually involve ratios of measurable quantities
which stem from similar underlying processes.

The WF law is universal across a wide range of
materials and temperature regimes because non-universal contributions due to,
e.g., density of states and effective mass  often cancel out.
Conversely, deviations from the WF law can be used to characterize the emergence
of new physical processes \cite{TanPPT07}.
Remarkably,  recent new experiments \cite{Tau15} show  that the WF law  holds  even at  quantum phase transitions.
Similarly, the Wilson ratio (WR) \cite{Som28,Wil75},
\begin{equation}
R_\mathrm{W}^{\chi}=\frac{4}{3}\left(\frac{\pi k_B}{\mu_B g_\mathrm{Lande} 
}\right)^2\frac{\chi}{c_V/T},
\label{eq-WR-chi}
\end{equation}
between the susceptibility $\chi$ and the specific heat $c_V$ divided by the
temperature $T$, measures the strength in magnetic fluctuations versus thermal
fluctuations \cite{Hew97,Sch99,Wan98}. Here $k_B$ is Boltzmann's constant, $\mu_B$ is the Bohr magneton
and $g_\mathrm{Lande}$ 
is the Lande factor.
This dimensionless ratio has been observed in a wide variety of Kondo systems \cite{Wil75,Hew97}.
Recent studies of the WR for magnetic states of a 1D spin ladder compound
\cite{Nin12} and the two-component attractive Fermi gas
\cite{GuaYFB13} show that $R_\mathrm{W}^{\chi}$ allows a convenient
identification of magnetic phases.
%
%
%
Dimensionless ratios therefore provide an elegant experimental and theoretical approach to a qualitative and quantitative characterization of the nature of complex multi-component quantum liquids.
%

\begin{figure}[tb]
{\includegraphics[width=0.95\columnwidth]
{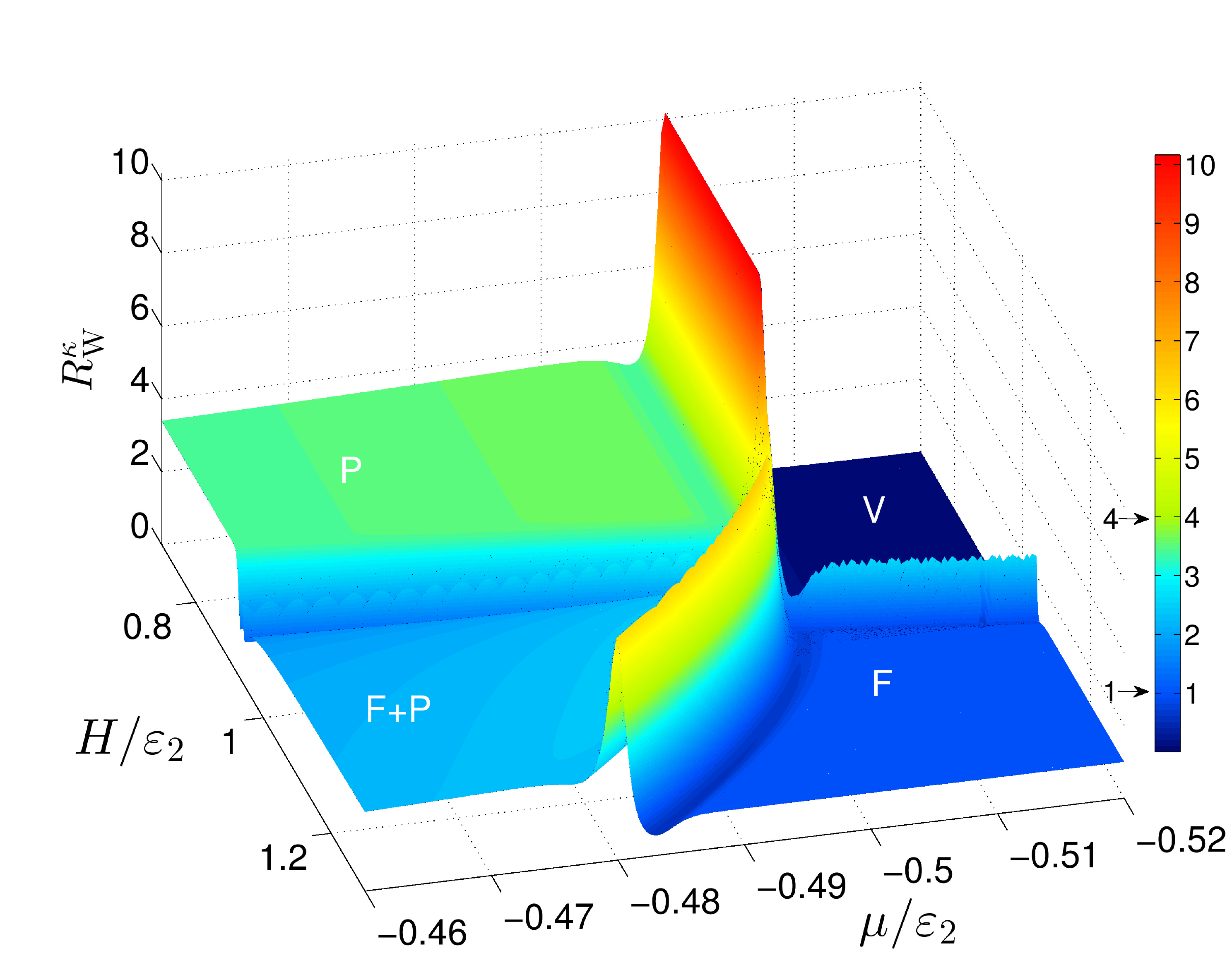}}
  \caption{
Phase diagram for the attractive SU(2) Yang-Gaudin  model at
$T=0.001\varepsilon_2/k_B$ given as 3D plot of the WR $R_{W}^{\kappa}$ in the
$\mu-H$ plane.
The pair binding energy is denoted by $\varepsilon_2=0.5$ and the interaction
strength $g_\mathrm{1D}=-2$. The expected values of $R_{W}^{\kappa}= 1, 4$ in the excess
fermion phase (F) and the pure pair phase (P) in strong coupling limit, respectively, are indicated on
the legend on the right. The mixed phase of fermions and pairs is denoted as F+P.
 }
\label{fig:WR-2c}
\end{figure}

In Fig.~\ref{fig:WR-2c}, we show that the \emph{Wilson-like} ratio
\begin{equation}
R_\mathrm{W}^{\kappa}= \frac{\pi^2 k_B^2}{3} \frac{\kappa}{c_V/T},
\label{eq-WR-kappa}
\end{equation}
relating particle fluctuations to energy  fluctuations, is even more successful
in determining phases in quantum liquids.
Recent experiment \cite{Yang-B:2016} show that  the  compressibility WR (\ref{eq-WR-kappa}) determines the  Luttinger parameter for the phase of TLL in 1D Bose gases and 
 characterizes the quantum fluctuations at quantum criticality.
The key observation underlying the predictive strength of both WRs is that the
phases are related to simple \emph{additivity rules} of the underlying elementary
excitations. In this paper, we shall prove  that such additivity rules are in fact
general for multi-component quantum liquids in 1D --- and we expect that they also hold for multicomponent Fermi liquids in higher dimensions \cite{Wan98}. 
  We further show  that both WRs
stem from these exact additivity rules, can quantitatively identify quantum phases of Tomonaga-Luttinger liquids (TLL)  independent of microscopic details and exhibit
universal scaling behavior at the quantum transitions between phases.
Their values within a given phase give information about the
internal degrees of freedom of a phase, e.g., its spin-degeneracy. Such
information is important, e.g., for the collective nature of multicomponent
interacting quantum liquids in cold atoms \cite{HoY99,WuHZ03,CazR14,Boett15},
large-symmetry fermionic systems \cite{Tai12,Sca14},  spin chains  \cite{Nin12,Kono15} and 1D Hubbard model \cite{Ess05}.

\section{Theory}

\subsection{General considerations}

In order to show the versatility of the WRs in identifying phases, we are
interested in systems that have a rich phase diagram. We therefore start 
by studying first an attractive quantum liquid that supports a
hierarchy of bound states. The system shall consist of unbound particles, pairs
of particles, triples of particles and so on until we have at most $w$-tuples.
For convenience, we shall denote a bound state formed from $r$ particles as a
\emph{$r$-complex}. Let $N_r$ denote the number of $r$-complexes which have
formed. Then the total number of particles is given as $N= N_1+  \ldots + w N_w$
and the density of a $r$-complex is $n_r=N_r/L$ such that the total
particle density is $n=\sum_{r=1}^w r n_r$.
The relative propensity of a phase is governed by a set of external fields
$H_r$, $r=1, \ldots, w$ with $H_w =0$. Here the $H_r$
are coupling to the (spin) moment of each $r$-complex. Then we can write for the
Hamiltonian of the system
\begin{equation}
{\cal H} = {\mathcal T} + {\mathcal V} - \sum_{r=1}^{w-1} H_r N_r - \mu N,
\label{eq-ham-general}
\end{equation}
where ${\cal T}$ and ${\cal V}$ are as of yet unspecified kinetic and many-body
interaction energies.
%

In order to compute the WRs \eqref{eq-WR-chi} and \eqref{eq-WR-kappa}, we need
to compute the susceptibility $\chi$, the compressibility $\kappa$ and specific
heat $c_V$ for the $w$-component system in the Gibbs ensemble $G(\mu, H_1, \ldots, H_w)$.
In a single compnent system, $\chi$ and $\kappa$ can be straightforwardly
computed as $\partial M/\partial H$ (with magnetization $M$) for constant particle number (canonical)
or $\partial n/\partial \mu$ for constant external fields (grand canonical). For a multi-component system, we
therefore define in complete analogy a chemical potential $\mu_r$ for each of
the $r$-complexes via
\begin{equation}
\mu_r=\mu+\frac{1}{r} H_r + \frac{\varepsilon_r}{r},
\label{eq-mur}
\end{equation}
where $\varepsilon_r$ denotes the binding energy of an $r$-complex.
Remarkably,  the  quantity $\mu_2$  has already been measured  in a recent experimental study of the equation of state for 2D  ultracold fermions \cite{Boett15}. It gives a deep  physical insight into the  crossover from Bose-Einstein condensate to Bardeen-Cooper-Schrieffer superconductor. 
Indeed, the choice \eqref{eq-mur} allows us to define a Fermi energy at $T=0$ for each fluid
of $r$-complexes in the same way that $\epsilon_F$ is defined in the
Landau's Fermi liquid  picture.
We now introduce a \emph{stiffness} in grand canonical and canonical  ensembles, respectively, as
\begin{eqnarray}
D_r^{\kappa}
&= &\frac{r }{\hbar\pi} \left(\frac{\partial \mu_r}{\partial n_r}\right)_{H_1, \ldots, H_{w-1}},\,\,D_{r,r'}^{\chi}
= \frac{r }{\hbar\pi} \left(\frac{\partial \mu_r}{\partial
n_{r^{'}}}\right)_{n}
\label{eq-stiffness-kappa}
\end{eqnarray}
for a $r$-complex fluid  subject to $\mu$ and  the field $H_{r'}$.
Also, the \emph{sound velocity} for $r$-complexes is defined as usual via
$v_r= \frac{d \epsilon_r (k)}{dk}|_{k_F}$,
where $\epsilon_r (k)$ is the dispersion relation for $r$-complexes and $k_F$ the Fermi momentum.
With these definitions, we can then prove in 1D that the individual $\kappa_r$, $\chi_r$
and $c_{V,r}$ in terms of the $D^{\kappa}_r$, $D^{\chi}_{r,r^{'}}$ and $v_r$ are given as
\begin{eqnarray}
\kappa_{r} & = & \frac{r^2}{ \pi\hbar} \frac{1}{D_{ r}^{\kappa}},\,\,
\chi_{r,r'} =  \frac{r^2}{ \pi\hbar} \frac{1}{D_{ r,r'}^{\chi}},\,\,
c_{V,r}          =  \frac{\pi k_B^2T}{3 \hbar} \frac{1}{v_r}
\label{eqs}
\end{eqnarray}
as shown in the Appendix.
We note that similar relations hold in Fermi liquids (FLs) \cite{Wan98}.
Our strategy therefore is to derive the densities $n_r$, the chemical potentials
$\mu_r$ and the dispersions $\epsilon_r(k)$ for the $r$-complexes as functions of
$H_1, \ldots, H_w$ and $\mu$. We expect that the definitions
\eqref{eq-stiffness-kappa} and \eqref{eqs}  are useful in general for interacting quantum liquids. However,
analytical or numerical access to these quantities is not necessarily
straightforward. Following a series of recent papers \cite{GuaBL13}, we prove here that integrable 1D
multi-component systems allow the explicit construction of $n_r$, $\mu_r$ and
$\epsilon_r$ using the thermodynamic Bethe Ansatz (TBA) \cite{GuaBLB07,Tak99,Sch93}.
Following earlier theoretical \cite{Wan98} and recent experimental \cite{Boett15,Yang-B:2016} findings, we furthermore speculate that similar relations also hold for FLs in higher dimensions and expect that such effective chemical potentials for multi-component systems can therefore
serve as convenient handles to describe multi-component quantum liquids.

\subsection{The $w$-component quantum liquid}

Let us now consider a specific $w$-component Hamiltonian \eqref{eq-ham-general} that supports multi-component quantum liquids. A convenient and experimentally
relevant example is the 1D SU$(w)$ Fermi gas with $\delta$-function interaction
confined to length $L$
\cite{Sut68,Tak70}. The system   is described by the   Hamiltonian (\ref{eq-ham-general})
with
\begin{eqnarray}
  \mathcal{T} + \mathcal{V}=-\frac{\hbar^{2}}{2m}\sum_{i=1}^{N}\frac{\partial^{2}}{\partial
x_{i}^{2}}+g_\mathrm{1D}\sum_{1\leq i<j\leq
N}\delta(x_{i}-x_{j}),\label{eq-ham}
\end{eqnarray}
and with the chemical potential $\mu$ and the  effective Zeeman energy $E_{z} =
\sum_{r=1}^{w}\frac{1}{2}r(w-r) n_{r}H_{ r}$.
%
There are $w$ possible hyperfine states $|1\rangle, |2\rangle, \ldots,
|w\rangle$ that the fermions can occupy.
Experimentally, $g_\mathrm{1D}$ ($=-2\hbar^2/m a_\mathrm{1D}$, with $a_\mathrm{1D}$ the effective scattering length in 1D \cite{Ols98}; from now on, we choose our units such that $\hbar^2=2m=1$ unless we particularly use the units.) can be tuned from a
weak interaction to a
strong coupling regime  via Feshbach resonances.
%
This  model  provides  an ideal experimental testing ground   to probe  few- and many-body physics \cite{Pag14,Lia10,WenMBL13}.

%
%
%


For the system described by \eqref{eq-ham}, the relation \eqref{eq-mur} follows naturally from the structure of the TBA
equations \cite{GuaBL13} (see Appendix \ref{App-TBA}). We will use in addition general energy-transfer relations for
breaking a $w$-complex into smaller $r$-complexes, i.e.
\begin{equation}
\frac{1}{r}H_{r} =  \mu_{r} - \mu_{w} + \frac{1}{w}\varepsilon_{w}-\frac{1}{r}\varepsilon_r 
\label{E-relation}
\end{equation}
with $ r=1,2,\ldots, w-1$.
Note that $\varepsilon_r=\frac{1}{48} r \left( r^2 -1\right) g_\mathrm{1D}^2$ is the explicit binding
energy for each $r$-complex in the system \eqref{eq-ham}.
Using the energy and particle conservation conditions, \eqref{E-relation} and $n=\sum_{r=1}^wrn_r$, respectively, we find that $\kappa$ and $\chi_r$ obey the additivity
rules,
\begin{eqnarray}
\kappa&=&\kappa_{ 1}+\kappa_{ 2}+\cdots +\kappa_{ w},\label{compressibility}\\
\frac{1}{\chi_{ r}}&=&\frac{1}{\chi_{r,1}}+\frac{1}{\chi_{ r,2}}+\cdots
+\frac{1}{\chi_{ r,w }}. \label{susceptibility}
\end{eqnarray}
in the TLL phases (see Appendix \ref{App-Wilson}). 
 Here the susceptibility $\chi_r$ represents the responses of different bound states to the change of the field $H_r$. Such
additivity appears naturally  for a non-interacting fluid.  
 Determining $v_{r}$ from the TBA, we similarly find that
\begin{equation}
c_{V}=c_{V,1}+c_{V,2}+\cdots +c_{V, w} \label{cv}.
\end{equation}

Based on the rules given in \eqref{compressibility}, \eqref{susceptibility} and
\eqref{cv}, we can construct the WRs of the $w$-component SU$(w)$ system to be
 \begin{eqnarray}
   R^{\chi }_{\mathrm{W},r'}
&= &
\left( \sum_{r=1}^{w} \frac{D_{r,r'}^{\chi}}{r^2}\right)^{-1}
\left( \sum_{r=1}^{w}\frac{1}{v_{r}} \right)^{-1},\label{WR1}\\
%
R^{\kappa}_\mathrm{W}
&= &
\left( \sum_{r=1}^{w} \frac{r^2}{D_{r}^{\kappa}} \right)
\left( \sum_{r=1}^{w} \frac{1}{v_{r}} \right)^{-1}. \label{WR2}
\end{eqnarray}
These ratios are dimensionless and uniquely determined by the sound velocities
and stiffnesses.
We note that the form of the WRs in (\ref{WR1}) and (\ref{WR2}) is similar for a $w$-component FL; all interaction effects have been included into \eqref{E-relation} via the choice of $\mu_r$.
%

In the strong coupling regime, $R^{\kappa}_\mathrm{W}$ for a pure $r$-complex phase ($R^{\chi}_\mathrm{W}=0$ in pure phases \cite{GuaYFB13}) can be
given in the form
\begin{eqnarray}
R^{\kappa,r}_{W}= r K_r =r^2 \left( 1-2B_r \frac{1}{|\gamma|}+B_r^2
\frac{1}{\gamma^2}\right) +O(\gamma^{-3}), \label{Rc}
\end{eqnarray}
where $ \gamma=g_\mathrm{1D}/2n$ is the dimensionless interaction strength,
$B_1=0$ and $B_r=\overset{r-1}{\underset{k=1}{\sum}} 1/k $ with $r=2,\ldots, w$.
In the above equation, $K_r$ is the phenomenological Luttinger parameter which can be directly  measured through the Wilson ratio $R^{\chi }_\mathrm{W}$.
This provides an important way to test the low energy  Luttinger theory. 
%
For $\gamma \to \infty,$ $R^{\kappa}_{W}$ then displays plateaus at the
integers $1^2, \ldots, r^2, \ldots, w^2$.
%
%
Deviations from integer $r^2= 1, 4, 9, É$ occur not because of a weakness of the ratios introduced here, but because of the physics underlying and determining the stability and mixture of bound states in a given phase. 
Thus $R^{\kappa}_{W}$ provides a convenient \emph{quantitative} phase
characteristic for quantum systems at finite $T$ as well as at $T=0$, see Fig.~\ref{fig:WR-2c} and below.

\subsection{Scaling of the WRs}

It is particularly interesting that the WRs identify non-TLL behaviour and quantum criticality in the quantum critical regime.
In fact, the WRs show sudden enhancement near a quantum phase transition  due to a breakdown of the quantum liquid nature, i.e.\ the vanishing linear dispersion for 1D systems.
For both WRs,  we have the scaling law 
\begin{eqnarray}
\hspace{-1.5em}
R^{\kappa,\chi}_\mathrm{W} &=&
\mathcal{F}^{\kappa,\chi} \left[ \frac{ ( \eta -
\eta_c)}{T^{\frac{1}{\nu z}}} \right]
+ \lambda_0 T^{-\beta} \mathcal{G}^{\kappa,\chi} \left[ \frac{ ( \eta -
\eta_c)}{T^{\frac{1}{\nu z}}} \right], \label{scaling}
\end{eqnarray}
where $\beta =(d/z)+1-2/(\nu
z)$ with $z=2$, $\nu=1/2$ and $d=1$ for 1D systems is universal; $\mathcal{F}^{\kappa,\chi}$,
$\mathcal{G}^{\kappa,\chi}$ are the scaling functions (see Appendix \ref{App-II}).
The second term in (\ref{scaling}) with $\beta =-1/2$ reflects a contribution
from the background at the temperatures above the energy gap $\Delta \sim | \eta -
\eta_c | ^{zv}$, with driving parameter $\eta$ (e.g.\ $\mu$ or $H_r$).
These critical exponents agree with the experimental result for the thermal and magnetic properties of the 1D spin chain at criticality \cite{Yang-B:2016,Kono15}.
For a phase transition from the vacuum to a TLL phase, $\lambda_0=0$.
Based on (\ref{scaling}), the slope of the temperature-rescaled
$R_{W}^{\kappa}$ curves at a critical point is given by $\left( \frac{\partial
R_{W}^{\kappa} }{\partial \mu} \right)_{\mu_c} = C_r /{T}$, see Appendix C.

The significance  of the WRs  can  also be understood from the quantum fluctuations of magnetization,
$\langle \delta M^2\rangle =\ell^d k_B T \chi$, and particle number, $\langle \delta N^2\rangle =\ell^d k_B T \kappa$; $\ell^d$ denotes the observation volume in $d$-dimensions.  Therefore, microscopically, $\chi$
and $\kappa$ measure the strength of these fluctuations just as $c_\mathrm{V}$ quantifies the energy  fluctuations; macroscopically, the temperature-independent compressibility and  the linear-temperature-dependent specific heat remarkably preserve the nature of quantum liquid at the renormalization fixed point, see Eq. (\ref{eqs}) like the Fermi liquid \cite{Sch99}. 
Consequently, $R^{\chi}_\mathrm{W}$ and $R^{\kappa}_\mathrm{W}$ characterise the
competition between fluctuations of different origin. A constant WR implies that
the two types of fluctuations are on an equal footing, regardless of the
microscopic details of the underlying many-body system. On the other hand, the growth of the WRs in the critical regime indicates the long-range character of the quantum fluctuations at the quantum phase transitions. 

\begin{figure}[tbp]
\vspace{-0.8cm}
\begin{center}
 {\includegraphics[width=0.95\columnwidth]
 {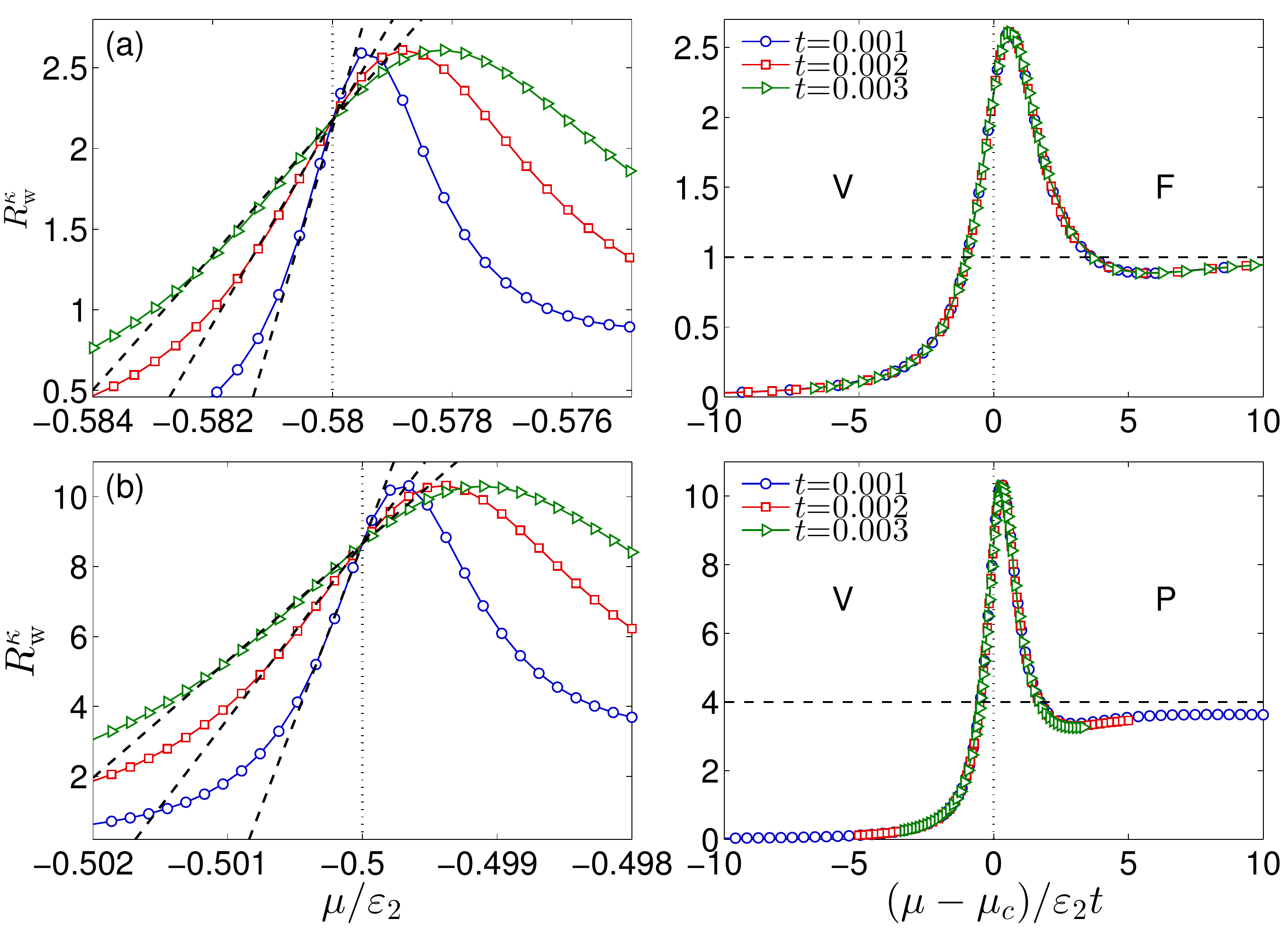}}
  \caption{
Scaling of $R^{\kappa}_{W}$ at phase boundaries (a) V-F and
(b) V-P with $H/\epsilon_2=1.16$ and $0.9$, respectively (cp.\ Fig.\
\ref{fig:WR-2c}).
The left panels in (a) and (b) show $R^{\kappa}_{W}$ vs.\ $\mu/\epsilon_2$ for
reduced temperatures $t=0.003$ ($\triangle$), $0.002$ ($\square$) and $0.001$
($\circ$). The vertical dotted lines indicate $\mu_c/\epsilon_2$.
The dashed lines show the calculated slopes $C_r/t$.
The right panels show the scaled $R^{\kappa}_{W}$ as in \eqref{scaling} with
$\beta=-1/2$ vs.\ $({\mu}-{\mu_c})/\epsilon_2 t$. The vertical dotted lines are
as in the left panel while the horizontal
lines indicate the strong interaction limit of $R^{\kappa}_{W}$ in the (a) F and
(b) P phases.
%
}
\label{fig-GY-scaling}
   \label{fig:boundary-Wc}
  \end{center}
\end{figure}

\section{Applications}

%
\subsection{The Yang-Gaudin model.} 

In order to show the usefulness of the new ratios, we start by considering the $w=2$
1D spin-$1/2$ Fermi gas with a $\delta$-function interaction \cite{Yan67,
Gau67}. This Yang-Gaudin  model has already provided an ideal experimental testing
ground \cite{Pag14,Lia10,WenMBL13} for few- and many-body physics
\cite{GuaBL13,Ors07,HuLD07}.
For $R_W^{\chi}$, the additivity rule of  susceptibility  and  its connection to the $D_r^{\chi}$'s following \eqref{WR1}  have been reported in \cite{GuaYFB13}.
For $R_W^{\kappa}$, we now find from \eqref{WR2} that
\begin{eqnarray}
 R^{\kappa}_{W}  &= &\left(
   \frac{1}{D^{\kappa}_{1}} + \frac{4}{D^{\kappa}_{2}} \right) / \left(
   \frac{1}{v_{1}} + \frac{1}{v_{2}} \right)\label{Rc-F2}.
\end{eqnarray}
Writing the explicit dependence on $\mu$, $H$ and $g_\mathrm{1D}$ is possible, but
tedious (we give explicit forms for $D^{\kappa}_{1,2}$ and $v_{1,2}$ in Appendix \ref{App-Wilson}).  In Fig.~\ref{fig:WR-2c} we show $R_W^{\kappa}$ in the $\mu$--$H$ plane.
$R_W^{\kappa}$ elegantly maps out the three finite temperature TLL phases of
$1$-complexes, i.e.\ fully-polarized fermions (F), $2$-complexes, i.e.\ pairs
(P), and the mixed TLL of excess fermions and pairs (F+P). The empty vacuum phase
(V) is also present in Fig.~\ref{fig:WR-2c}.
In particular, we note that for the pure phases F and P, $R^{\kappa}_{W}  \approx r^2$, i.e.\ close to the strong coupling value $r K_r$ 
\cite{CapRLA08}.
%
%
Hence for the Yang-Gaudin model in Fig.~\ref{fig:WR-2c} we see $R_\mathrm{W}^{\kappa}=1$ in phase F and $4$ in phase
P. In mixed phases, e.g.\ F+P, such a constant plateau no longer exists.
For $R_W^{\kappa}$ in the critical regime, we find a rapid increase due to a
strong increase in thermal fluctuations.
In Fig.\ \ref{fig:boundary-Wc} we show the scaling behavior \eqref{scaling} for
$R_W^{\kappa}$ close to the transition from the vacuum phase into the F and the
P phases. We find $ C_1 \approx 1.2546$ and $C_2 \approx 2\times 5.0185$, respectively.
Eq.\ \eqref{scaling} does not contain any free fitting parameter, so one can use this scaling to determine the temperature of the quantum
liquid.
The scaling law is also similar to the scaling in the non-FL regime of heavy fermions \cite{Hew97,GegSS08}.

\begin{figure}[tb]
\vspace{-0.7cm}
{\includegraphics[width=0.95\columnwidth]{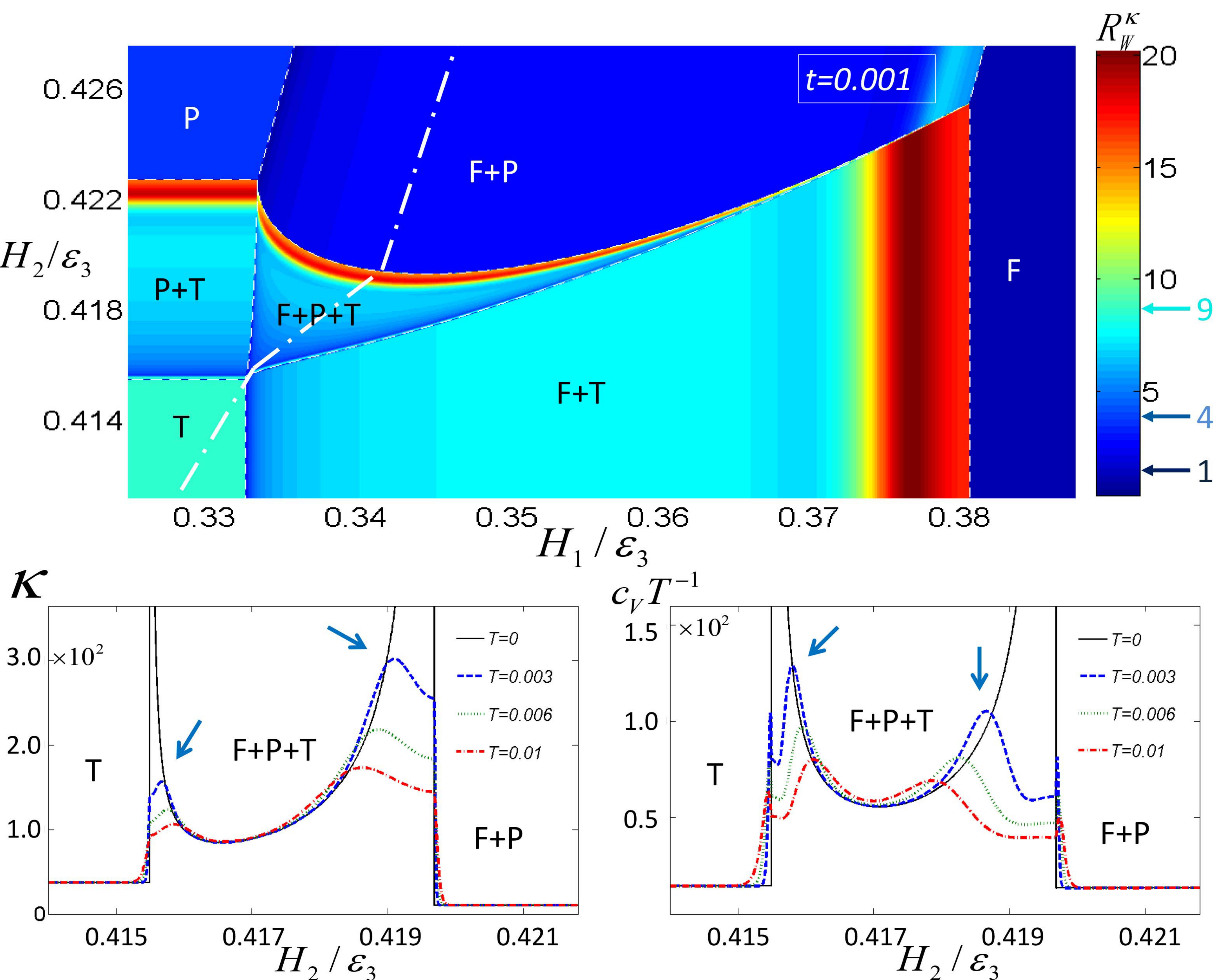}}
\caption{Upper panel: plot of $R_{W}^{\kappa}$ for the 1D
three-component attractive Fermi gas in the $H_1$--$H_2$ plane at temperature
$t=0.001 \varepsilon_3/k_{B}$.
  %
 %
 %
The values of $ R_{W}^{\kappa}= 1, 4, 9$ in the pure phases F, P and T,
respectively, are indicated by the colors marked on the color scale shown.
The dashed lines follow the phase boundaries as indicated.
Lower panel: compressibility $\kappa$ and specific heat $c_V/T$ vs.\ external
field $H_2/\varepsilon_3$ for a fixed choice of polarization $n_1=n_2$ indicated by
the dash-dotted line in the phase diagram (upper panel).
At zero temperature, $\kappa$ and $c_V/T$ satisfy the additivity rules
(\ref{compressibility}) and (\ref{cv}) as shown by the solid lines while at
finite $T$, they exhibit peaks as indicated by the arrows for $t=0.003\varepsilon_3/k_B$.
%
Near the phase boundaries both $\kappa$ and $c_V/T$ diverge quickly as
$(H_2-H_2^c)^{-1/2}$.
}
  \label{fig:Wc-3}
\end{figure}

\begin{figure}[tb]
\vspace{-0.5cm}
{\includegraphics[width=0.95\columnwidth]{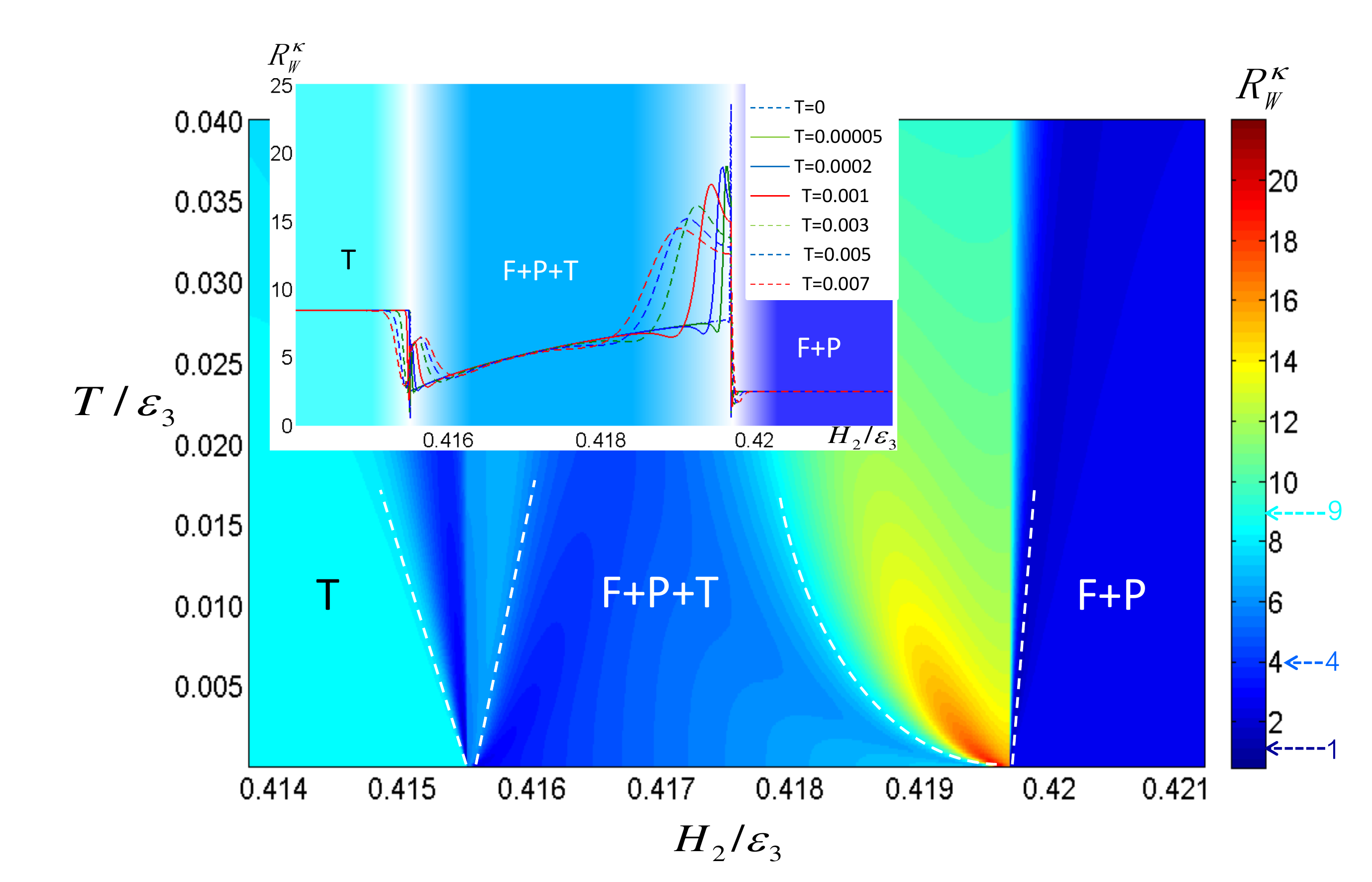}}
  \caption{Mapping out quantum criticality with  $R_{W}^{\kappa}$  in the $T$--$H_2$ plane. The $H_2$ values have been chosen to follow the fixed polarization $n_1=n_2$ as indicated by the dash-dotted white line in Fig.~\ref{fig:Wc-3}.
  %
  %
  %
  The white dashed lines denote the crossover temperatures $T^{*}$ beyond which the TLL phases break down.
  The inset shows individual temperature curves of $R_{W}^{\kappa}$ vs.\ $H_2$ for representative temperatures.
  The bold solid line shows the result corresponding to (\ref{WR2}) while the thin dashed lines show the numerical results obtained from the TBA equations (Appendix \ref{App-TBA}).
  %
  %
    }
  \label{fig:Wc-3-mu}
\end{figure}


\subsection{A $w\ge3$ quantum liquid.} 

An even richer phase diagram exists for the three-component $\delta$-function
interacting Fermi gas with an attractive interaction \cite{Sut68}. Its
$T=0$ phases are known to consist of excess fermions (F), pairs (P), trions (T)
and mixtures thereof \cite{GuaBLZ08,KuhF12}.
Using the TBA equations, we again numerically and analytically
calculate the WRs and determine their $H_1$, $H_2$ and $T$ dependencies. In Fig.~\ref{fig:Wc-3} we show that in the pure
phases F, P and T, we have $R_{W}^{\kappa}= 1, 4, 9$, respectively, and the
phase boundaries are clearly marked by large increases in $R_{W}^{\kappa}$ near
the critical points.
At $T>0$, we see that the $\kappa$ and $c_V$ curves become progressively more
rounded across the phase transitions for $T$ increases. The magnetic field
associated with the position of the finite-height peaks varies as a function of
temperature. This can be used to define a crossover temperature $T^{*}$ for each
such magnetic field corresponding to a peak.
For $T<T^{*}$, our analytical and numerical results reconfirm the validity of the additivity rules  (\ref{compressibility})--(\ref{cv}), i.e.\  $\kappa$ and $\chi$  remain independent of the temperature  and $c_V$ depends linearly on the temperature in the TLL phase (cp.\ lower panel in Fig.~\ref{fig:Wc-3}).
This nature is also seen in Fermi liquids \cite{Wil75,Sch99}.
The additivity nature of the susceptibility is presented in Appendix \ref{App-Wilson}.
The significance of such additivity rules is the characteristic of quantum liquids at the renormalization fixed point, such as TLLs in integrable models and Fermi liquids in metals.  etc. 
For $T>T^{*}$, $R_{W}^{\kappa}$ exhibits the critical scaling behaviour (\ref{scaling}) (see also Appendix \ref{App-II}).
Fig.~\ref{fig:Wc-3-mu} shows the contour plot of $R_{W}^{\kappa}$ at  low temperatures.  The universal quantum critical behaviour (\ref{scaling}) is characterized by the exponents $z=2, \nu=1/2$. For $T<T^{*}$, the critical exponents are $z=\nu=1$.

 Moreover, using the TBA equations for the $SU(w)$ Fermi gases with an attractive interaction  \cite{LeeGBY11}, we calculate the susceptibility $\chi_r =\partial M /\partial H_r$  in response to the change of magnetic field $H_r$ for the \emph{spin-gapped} phases. The other magnetic fields are fixed and the magnetization is given by $M=\sum_{r=1}^{w-1}n_r r(w-r)/2$. In dimensionless units,  the energies are rescaled by the interaction energy $\epsilon_b=\frac{1}{2}\hbar^2 c^2/(2m)$, i.e.\ the temperature $t=T/\epsilon_b$, the pressures $\tilde{p}^r=p^r/(|c|\epsilon_b)$, the susceptibility $\tilde{\chi}=|c|\xi/2$, see the Appendix B. We  obtain explicitly a general expression of the susceptibility $\chi_r$ for the gapped phase in which a small number of  the $r-$complexes    are created due to the change of $H_r$  for $t\ll1 $,
\begin{equation}
\tilde{\chi}_r\approx \frac{1}{4\sqrt{2\pi}}\frac{1}{\sqrt{t}}\sqrt{r}(w-r)re^{-\frac{\Delta_r }{t}}, \label{chi-gap}
\end{equation}
where the gap is given by 
\begin{equation}
\Delta_r\approx -r\tilde{\mu}_r +\sum_{m=1}^w\sum^{\tiny{\mathrm{min}(r,m)}}_{\tiny \begin{array}{c}q=1\\
2q\ne r+m \end{array} }\frac{4\tilde{p}^{m}}{m(r+m-2q)},
\end{equation}
see the Appendix C. 
For example, from  the upper panel phase diagram in Fig. \ref{fig:Wc-3}, we observe  that in  the gapped phase of T, 
$\tilde{\chi}_1$ with $\Delta_1=-\tilde{\mu}_1+2\tilde{p}^3/3$ and $\tilde{\chi }_2$ with $\Delta_2=-2\tilde{\mu}_2+16\tilde{p}^3/9$ show a dilute magnon behaviour related to the phase transitions from the phase T into the mixed phase F+T and into the  the mixed P+T,  respectively. They decay exponentially as the temperature approaches zero.  By properly choosing $H_1$ and $H_2$, we also can have a phase transition from the gapped phase T into the mixed phase F+P+T.


\subsection{Critical theory for the $SU(w)$ repulsive Fermi gases}


%
The existence of internal degrees of freedom significantly changes the
quantum magnetism and the dynamics of the system compared to the spinless Bose
gas.
It is well established that the critical behaviour of the  spin $SU(w)$ chain can be described by the Wess-Zumino-Witten $\sigma$ model of level $\ell=1$ with the  Kac-Moody central charge 
$C_s=\ell(w^2-1)/(\ell +w)$ \cite{Affleck:1987}. The $w$-component repulsive Fermi gases  display a $U(1)\otimes SU(w)$ symmetry characterized by one charge degree of freedom and $w-1$ spin rapidities. 
The low-energy physics of the system  is described by the spin-charge separated conformal field theories of an effective Tomonaga-Luttinger liquid and an antiferromagnetic $SU(w)$ Heisenberg spin chain \cite{Ess05,Frahm:1990}. By using the TBA equations given in \cite{LeeGBY11} with $H\to 0$ we find that the pressure of the 1D repulsive Fermi gases is given by 
\begin{equation}
p=p_{T=0}+\frac{w(w^2-1)H^2}{24\pi v_{\rm s}}+ \frac{\pi T^2}{6} \Big[\frac{1}{v_{c}}+\frac{C_s}{v_{s}}\Big],
\end{equation}
where $p_{T=0}$  is the pressure at $T=0$ and $v_{s,c}$
are the pseudo Fermi velocities in  the spin  and charge  sectors, respectively.
This result is consistent with the critical field theory for the $SU(w)$ spin chains \cite{Affleck:1987}.
In the zero magnetic field limit, the spin and charge velocities for the Fermi
gas with strong repulsion are, respectively,
\begin{eqnarray}
v_{s}&=&\frac{2 n^2\pi^3a_\mathrm{1D}}{3w} \left(1+3a_\mathrm{1D}Z\, n\right), \nonumber\\
v_{c} &= &2n\pi \left(1+2a_\mathrm{1D}Z\, n\right),\label{velocity-SC}
\end{eqnarray}
where $Z=-\frac{1}{w}\left[ \psi (\frac{1}{w})+C \right]$ and $C$ the Euler
$\gamma$-constant, $\psi(x)$ the Euler $\psi$-function.
The susceptibility is then given by Luttinger-liquid relation 
\begin{equation}
\chi v_s=\frac{1}{12\pi }w(w^2-1)
\end{equation} 
 in the limit $H\to 0$.
Indeed the sound velocity $v_c$ (\ref{velocity-SC}) of the Fermi gas in the
large-$w$ limit coincides with that for the spinless Bose gas, whereas $v_s$
vanishes quickly as $w$ grows.
This gives a reason why the quantum liquid of multi-component fermions reduces to the liquid of a spinless Bose gas in this limit \cite{Pag14}.
In this sense, $R_{W}^{\chi}$ captures this unique large-spin charge separation
mechanism in the $w$-component repulsive Fermi gas.
Its explicit expressions is 
\begin{eqnarray}
R_{W}^{\chi}&=&\frac{w(w^2-1)v_c}{3\left[ (w-1) v_c +v_s\right]},\label{WR-SC}
\end{eqnarray}
displaying plateaus of height $w(w+1)/3$ for either strong repulsion or in the large-$w$ limit, hence capturing the spin degeneracy. For example, $R_W^{\chi}=2$ and $4$ for the two- and three-component Fermi gases with strong repulsion, respectively.
\begin{figure}[tb]

\begin{center}
\includegraphics[width=0.95\columnwidth]
{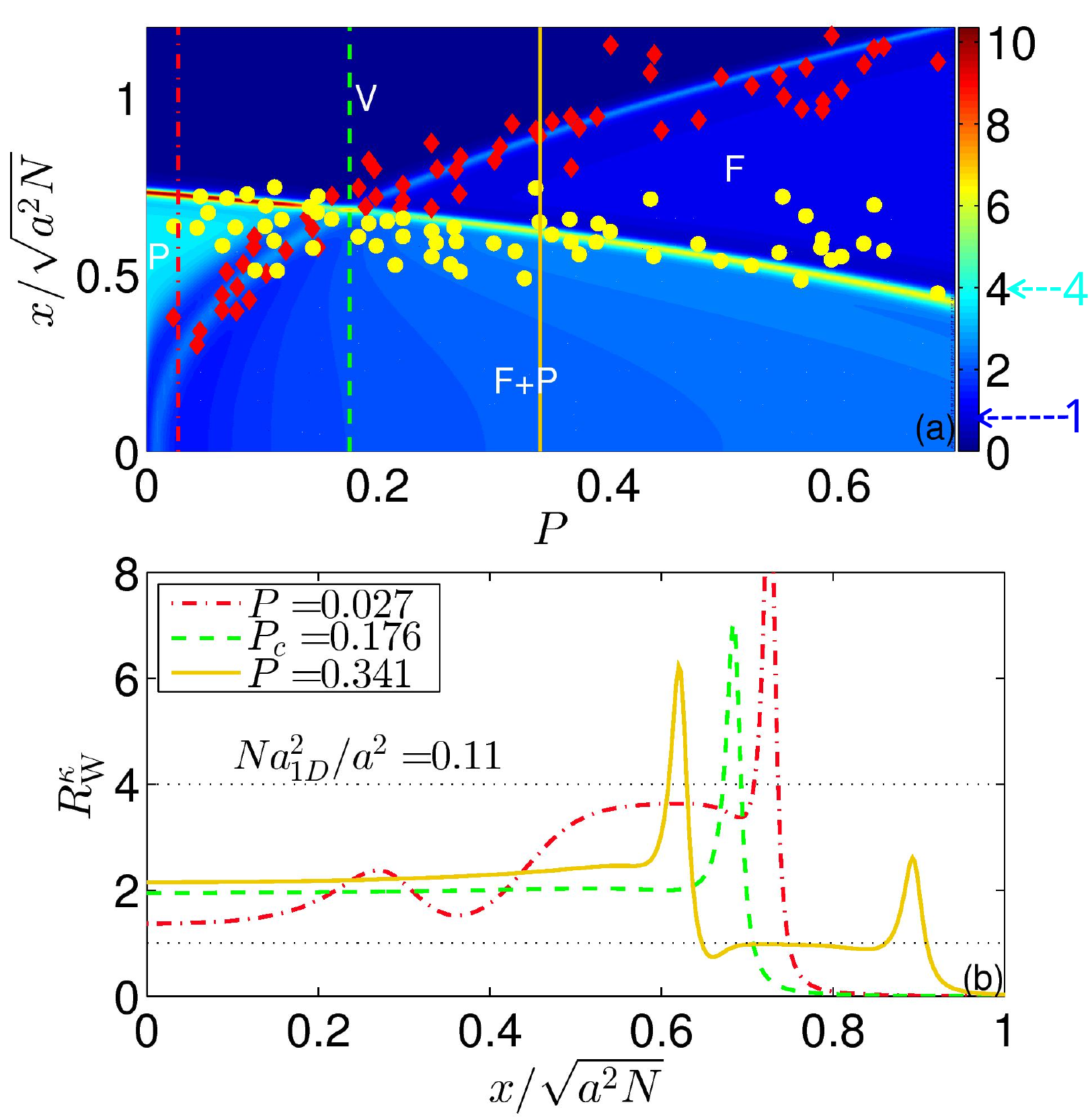}
\caption{Upper panel: 
Contour plot of the WR $R_{W}^{\kappa}$ (\ref{eq-WR-kappa}) for the Yang-Gaudin model in a harmonic trap with fixed
particle number $N$ and three different polarizations $P=n_1/n$ at the temperature $T=
0.001\varepsilon_2/k_B$.
The WR evolves thin round peaks near the phase boundaries, whereas the values of the WR (\ref{Rc-F2}) quantify different quantum phases of the TLLs. 
The WR peaks  are in good agreement with the experimental phase boundaries (red and yellow dots) observed in \cite{Lia10}.
Lower panel:  the red-dashed and yellow-solid  lines show  the WR $R_{W}^{\kappa}$ in 
the trapped gas with the polarizations $P=0.027$ and $P=0.341$, respectively. 
They indicate  segments of the Fulde-Ferrell-Larkin-Ovchinnikov (FFLO) phase in the centre accompanied by the
wings of the P  and the F states, respectively.
The highest $R_{W}^{\kappa}$ plateau approaches $4$ suggesting a quasi-1D
superfluid nature, whereas the lowest plateau shows the free fermion nature with
$R_{W}^{\kappa}=1$.
For a critical polarization $P_c = 0.176$, the green-dashed line shows that
the trapped gas consists solely of the FFLO-like state.
%
%
}
\label{fig:WR-trap2}
\end{center}
\end{figure}

\section{Experimental realizations}

Both WRs $R_{W}^{\chi}$ and $R_{W}^{\kappa}$ are readily accessible by experiments. 
For example, the 1D SU$(w)$ $\delta$-function interacting Fermi gas of ${}^{173}$Yb
has been realized experimentally \cite{Pag14}.
It was shown that in the large-$w$ limit the ground state of the gas with a
repulsive interaction exhibits properties of a bosonic spinless liquid.
 In the context of ultracold atoms, it is highly desirable to measure the quantum criticality  and the TLL in such Fermi gases with rich spin and charge degrees of freedom. 
In this scenario, the Yang-Gaudin model is an ideal model to conceive critical phenomena induced from spin and charge interaction effects. This model was recently  studied via an ultracold atomic gas in a harmonic trap, such as the two-component ultracold ${}^6$Li atoms of Ref.~\cite{Lia10}.  
Due to the harmonic confinement,
the chemical potential in the equation of state
should be replaced by $ \mu \left( x \right) = \mu \left( 0 \right) - m \omega^2
x^2/ 2$ (within the local density
approximation). Here $x$ denotes the position along the 1D trap. Changing $x$ is then equivalent to changing $\mu$ and different phases are located at different spatial positions along the trap.
Using a rescaled coordinate $y=x a_\mathrm{1D}/ (2a^2)$ with axial characteristic oscillation
length $a=\sqrt{\hbar /m\omega }$, the density profile of the trapped gas can be
determined from the dimensionless quantities $Na_\mathrm{1D}^2/a^2=2a_\mathrm{1D} \int_{-\infty}^{\infty} dy\,n(y)
$ and polarization $Ma_{\rm
1D}^2/a^2=2a_\mathrm{1D} \int_{-\infty}^{\infty} dy \,m_z(y) $.
The compressibility can be extracted via $ \kappa =\partial n /\partial \mu
=-\partial n/\partial x / (x m\omega^2)$ \cite{CheY07,HoZ10},
while the density $n(x)$ is read off from experimental profiles along $x$.
%
The specific heat
can be measured through the sound velocities,
while the density waves of pairs and unpaired fermions can be experimentally
created by a density depletion or pulse with a far-detuned laser beam
\cite{And97}.
 In the upper  panel of the Fig.~\ref{fig:WR-trap2}, we observe that $R_{W}^{\kappa}$  at a very low temperature naturally maps out the phase diagram of the system, showing a good agreement with the experimental phase boundaries \cite{Lia10}. 
In the lower panel of Fig.~\ref{fig:WR-trap2}, we show the behaviour for $R_{W}^{\kappa}$,
exhibiting the plateau structure as the system passes through different phases
upon changing $x$.

\section{Conclusions}
In summary, we have shown that the additivity rules of physical properties  reveal an  important  characteristic   of the TLL and  provide the physical origin  of the dimensionless WRs.
%
Such  dimensionless  ratios can be used to identify full  TLL phases and capture  the essence of   phenomena ranging  from  quantum criticality to  spin and charge separation in  a wide variety of 1D  many-body systems.
 We  have presented some universal relations, such as  the WRs (\ref{WR1}) and (\ref{WR2}),  the Luttinger parameter (\ref{Rc}), the dimensionless scaling function of the WR (\ref{scaling}), the susceptibility for the gapped systems (\ref{chi-gap}) and the WR of $SU(w)$ repulsive Fermi gas related to the  level-$1$  Wess-Zumino-Novikov-Witten conformal theory (\ref{WR-SC}). 
We also show excellent agreement between the phase diagram predicted by the WRs and the experimentally determined one for the Yang-Gaudin model \cite{Lia10}. 
Our results therefore successfully demonstrate how to predict universal laws for experimentally realizable quantum liquids in  Bose and Fermi degenerate gases,  Bose-Fermi mixtures, 1D Hubbard model, strongly correlated electronic systems, spin compounds  near to and far from quantum critical points on an equal footing.


\acknowledgments
We thank T.\ Guttmann, J.\ Ho, R.\ Hulet, G.\ Shlyapnikov, Z.-C.\ Yan,
S.-Z.\ Zhang, Q.\ Zhou and H.\ Zhai for helpful discussions.
This work has been supported by the National Basic Research Program of China
under Grant No. 2012CB922101 and No. 2011CB922200, by the key NNSFC grant No. 11534014 and by the NNSFC under grant
numbers 11374331 and 11304357.
XWG thanks the Department of Physics at Rice University, University of
Washington and ITAMP at Harvard University for kind hospitality.
RAR gratefully acknowledges the hospitality at WIPM, CAS Wuhan and funding via a CAS Senior Visiting Professorship.




\setcounter{figure}{0}
\def\thefigure{A\arabic{figure}}

\appendix

\section{Thermodynamic Bethe Ansatz for the SU($w$) Fermi gas}
\label{App-TBA}

The Gibbs free energy of the SU($w$) attractive gas is 
\begin{equation}
 G({\mu}, H_1, \ldots,H_{w})
  =
  \sum_{r=1}^{w} \frac{r T}{2 \pi} \int d k \ln \left[1 +  e^{- \epsilon_{r} (k) / T}\right]\label{TBA-G}
\end{equation}
with dispersions $\epsilon_{r}(k)$ defined in the TBA via 
\begin{widetext}
\begin{eqnarray}
\epsilon_{r}(k)
&=&rk^2 - r{\mu}- H_r - \varepsilon_r
+ \sum_{p=1}^{r-1} \left\{ \sum_{q=p}^{w} a_{q + r - 2 p} \ast T \ln \left[1 + e^{-\epsilon_{q}(k) / T} \right] 
+ \sum_{q=r+1}^{w} a_{q - r} \ast T \ln \left[1 + e^{- \epsilon_{q}(k) / T}\right] \right\} \nonumber \\
&& - \sum_{q=1}^{\infty} a_q \ast T \ln \left[1 + e^{- \eta_{r,q} / T}\right],\\
\eta_{r,l}(k) &=& l \cdot \left(2 H_r - H_{r - 1} - H_{r + 1} \right) 
+ a_l \ast T \ln \left[1 + e^{- \epsilon_r(k) / T}\right] 
+ \sum_{q=1}^{\infty} U_{lq} \ast T \ln \left[ 1 + e^{- \eta_{r,q}(k) / T} \right] \nonumber \\
&& - \sum_{q=1}^{} S_{lq} \ast T \ln \left[1 + e^{- \eta_{r - 1,q}(k)/ T}\right] 
- \sum_{q=1}^{\infty} S_{lq} \ast T \ln \left[1 + e^{- \eta_{r + 1,q}(k) / T} \right],
\end{eqnarray}
with
\begin{eqnarray}
  {U}_{lj}(x) &= &
  \left\{ 
  	\begin{array}{ll}
	a_{| l - j |} (x) + 2 a_{| l - j | + 2} (x) + \ldots + 2 a_{l + j - 2}(x) + a_{l + j} (x), & l \neq j\\
	2 a_2 (x) + 2 a_4 (x) + \ldots + 2 a_{2 l - 2} (x) + a_{2 l}(x), & l=j, 
 \end{array}\right.\nonumber\\
S_{lj}(x)& = &
  \left\{ 
  	\begin{array}{ll}
	a_{| l - j | + 1} (x) + 2 a_{| l - j | + 3} (x) + \ldots + 2 a_{l + j -     3} (x) + a_{l + j - 1}(x), & l \neq j\\
	a_1 (x) + a_3 (x) + \ldots + a_{2 l - 3} (x) + a_{2 l - 1} (x), & l=j ,
 \end{array}\right.\nonumber 
\end{eqnarray}
\end{widetext}
while $\ast$ denotes the convolution
$
 (a \ast b) (x) = \int a(x-y) b(y) dy $
and the $\eta_{r,l}(k)$ represent the spin string parameters; furthermore
$a_n (x) = {n | c |}/2 \pi[(n c / 2)^2 + x^2]$. 
Here $c=mg_{\rm 1D}/\hbar^2=-2/a_{\rm 1D}$.
The numerical results used in the figures were obtained  by solving the above TBA equations. 

\section{Additivity rules for SU($w$)  Fermi gases}
\label{App-Wilson}

From \eqref{eq-mur}, we find that constant fields in the grand canonical ensemble, i.e.\ $d H_{1} =d H_{2} = \ldots =d H_{w-1} =0$, imply $d\mu_r = d\mu$ for $r=1, \ldots, w$. Consequently, we have
\begin{equation}
\kappa 
= \frac{\partial n}{\partial \mu} 
=  \frac{\partial \sum_{r=1}^{w} r n_r}{\partial \mu} 
= \sum_{r-1}^{w}  \frac{r \partial n_r}{\partial \mu_r} 
= \sum_{r=1}^w \kappa_r ,      
\end{equation}
with $\kappa_r$ as given by \eqref{eq-stiffness-kappa} and \eqref{eqs}. Note that this result is general and does not use special properties of the SU($w$) gases.

The additivity rules for the spin susceptibility are very intriguing. For our convenience in analysis of  the SU(w) case,  we prefer to  keep the ratios among the densities of the charged bound states as 
$n_1:n_2:\,\ldots \,:n_{w-1}=\lambda_1:\lambda_2:\,\ldots\,:\lambda_{w-1}$,  then we can parameterize the $n_1,n_2,\ldots ,n_w$ as
\begin{eqnarray}
 n_{r} =  \frac{\lambda_r}{\lambda}   \hat{n} , (r=1,2,\ldots,w-1),\, \,n_{w}  =  \frac{1}{w}   (  n- \hat{n} )\label{n_r-ratio}
\end{eqnarray}
where $\lambda=\lambda_1+2 \lambda_2+...+(w-1) \lambda_{w-1}$.
In order to compute the additivity rules for $\chi_r^{-1}= \partial H_r / \partial M$, with magnetization $M= \sum_{r=1}^{w-1} r(w-r)n_r/2$, we start with the Legendre transformation from Gibbs free energy to ground state energy:
$E=G+\mu n +\sum_{r=1}^{w-1} n_r H_r $, and in the ensemble $\{n_1,\cdots,n_{w-1},n \}$ the field $H_r$ can be obtained:
\begin{align}
&H_r = \frac{\partial E}{\partial n_r} \Big{|}_{n_1,\cdots,n_{r-1},n_{r+1},\cdots,n_{w-1},n}  \notag \\
&=\frac{\partial E}{\partial n_r} \Big{|}_{n_1,\cdots,n_{r-1},n_{r+1},\cdots,n_{w}}
-\frac{r}{w}\frac{\partial E}{\partial n_r} \Big{|}_{n_1,\cdots,n_{w-1}}\label{H_r}
\end{align}
where $E= \sum_{r=1}^{w} E_r$ denotes the energy of the multi-component ground state \cite{GuaBL13} and we have $\partial E / \partial n_r= r \mu_r$ for consistency with \eqref{eq-mur}. 
We emphasize that the additivity of $E$ is a fundamental property of a TLL, implied by the linearity of the dispersions \cite{GuaBL13}. 

Following the relation (\ref{n_r-ratio}), we can define  differential forms of any thermodynamic function $f=f(n_1,
\cdots, n_w)$ :
\begin{eqnarray}
     \textmd{d} \, f &=&  \underset{r=1}{\overset{w-1}{\sum}} \frac{\partial f}{\partial n_r} \cdot
     \frac{\lambda_r}{\lambda} \Delta \hat{n}- \frac{\partial  f}{\partial  n_{w}} \frac{1}{w} \Delta
     \hat{n}_{},    \\
     \textmd{d}\,  M &=& \underset{r=1}{\overset{w-1}{\sum}} \frac{1}{2} r(w-r)
     \frac{\lambda_r}{\lambda} \cdot \Delta \hat{n}.
\end{eqnarray}
We denote the operator $\mathcal{D}_r= \frac{\partial}{\partial n_{r}} -  \frac{r}{w}
     \cdot \frac{\partial}{\partial  n_{w}} $ with $r=1,2,\ldots, w-1$.
Then the field $H_r$ can be expressed explicitly: $H_r=\mathcal{D}_r E$
 and  the susceptibility  in response  to the $H_r$ can be expressed as
\begin{equation}
 \frac{1}{\chi_{r}} = \frac{\texttt{d}\,   H_{r}}{\texttt{d}\,   M} \nonumber   
 = \frac{\sum_{r=1}^{w-1} \lambda_r\mathcal{D}_r}{ \sum_{r=1}^{w-1} \frac{1}{2} r(w-r) \lambda_{r} }\cdot \mathcal{D}_{r}  E. \label{ch-r}
 \end{equation}
 We note that the last term in the Eq. (\ref{H_r}) linearly depends on  the density $n_r$ and $n_w$. Therefore it can be safely dropped off in the calculation of the susceptibilities because of the second order of derivatives. 
In terms of the ground state energies $E=\sum_{\ell =1}^wE_{\ell}$  for the individual charge bound states 
We define the stiffness as
\begin{eqnarray}
  \frac{1}{\chi_{ r\ell}} = \frac{\hbar\pi}{r^2} D^\chi_{r,l}  = 
  \frac{\sum_{r=1}^{w-1} \lambda_r\mathcal{D}_r}{ \sum_{r=1}^{w-1} \frac{1}{2} r(w-r) \lambda_{r} }\cdot \mathcal{D}_{r}  E_{\ell},
  \end{eqnarray}
where $r=1,2,\ldots, w-1$ and $\ell=1,2,\ldots, w$.
Consequently,  we have the susceptibility (\ref{susceptibility}) in responses to the external field $H_{r}$.

\begin{figure}[htp]
\begin{center}
	 \includegraphics[width=1.0\linewidth]{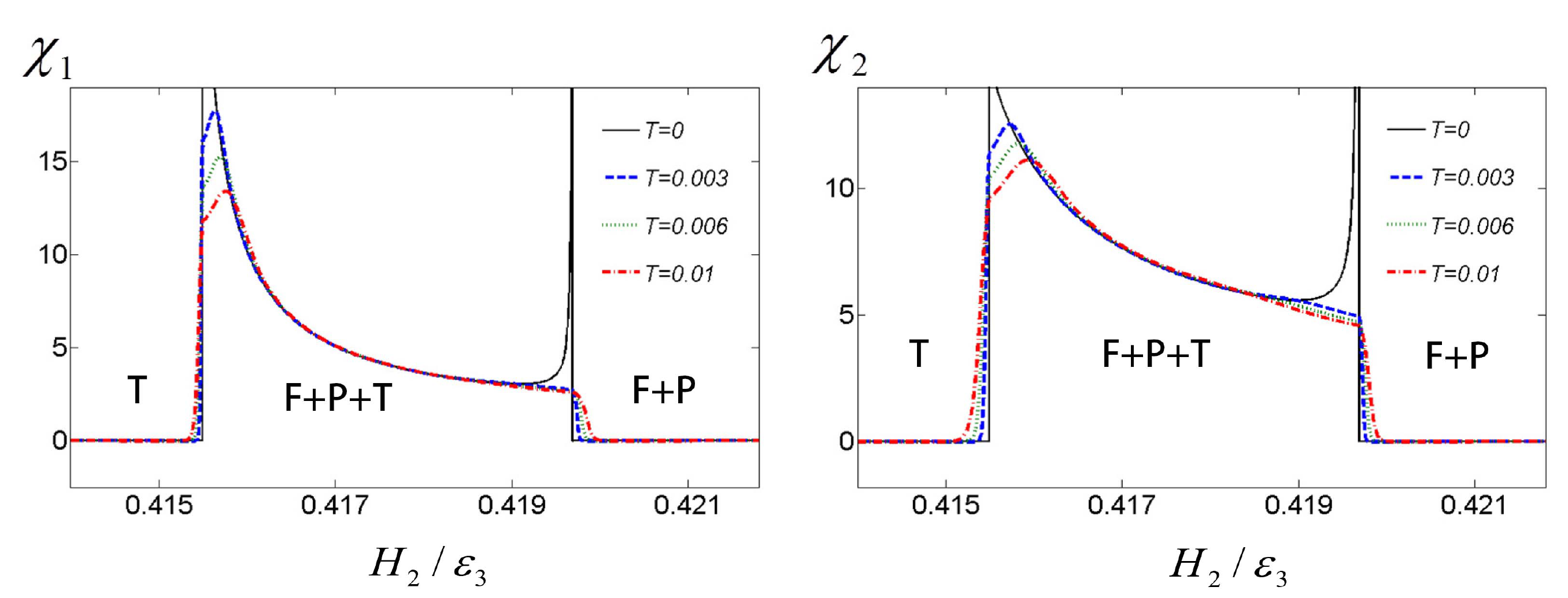}
	\caption{\footnotesize{ Susceptibilities in response  to $H_1$ and $H_2$ for the three-component Fermi gas with a fixed ratio of $n_1/n_2=1$ at finite temperatures.
Under this setting the  $H_1$  and $H_2$  have a one-to-one mapping along the line of $n_1/n_2=1$ with a fixed total density.
 These two  figures show that the susceptibility is temperature-independent for the TLL phase of F+P+T indicated by the long dashed line in Fig. \ref{fig:Wc-3}  in the main text.
 The deviation from the solid lines in the two figures show a breakdown of the TLL.  Here T, B and F stand for trions, pairs and excess fermions, respectively.
 The solid lines confirm the additivity rule Eq.  (\ref{susceptibility})  in the main text.
}}
	\label{sus-H2}
\end{center}
\end{figure}

 In fact, Zeeman spliting can be characterized by the Zeeman energy levels $\epsilon_Z^r$ or by the effective magnetic fields 
$H_r$ with $r=1,2,\ldots, w$. Here $H_w=0$.  Both sets of parameters
are related via the relation
\begin{equation}
\sum_{r=1}^w\epsilon_Z^r n^r = -\sum_{r=1}^w H_r (n^r - n^{r + 1}).
\end{equation}
A consistent solution of this equation gives the relations between $H_r$ and Zeeman energy levels \cite{LeeGBY11}.
 If we denote the difference between the energy
levels of fermions in the states $| m  + 1 \rangle$ and
$| m  \rangle $ as $\Delta_{m + 1, m} = \epsilon_Z^{m + 1}
- \epsilon_Z^m$,
then the total susceptibility $\chi = \frac{\texttt{d}\, M}{\texttt{d}\, \Delta_{ \rm total}}$ with $\Delta_{ \rm total}=\sum_{r=1}^{w-1} \Delta_{r+1, r}$  is given by 
\begin{eqnarray}
 \chi^{- 1} &=& \frac{\texttt{d}\,  \sum_{r=1}^{w-1} \Delta_{r+1, r} }{\texttt{d}\, M} 
 = \frac{\texttt{d}\,( H_1
   + H_{w - 1})}{\texttt{d}\, M}\nonumber\\
   && = \frac{1}{\chi_1} + \frac{1}{\chi_{w - 1}}.
   \end{eqnarray}
Using the numerical calculation  for the Gibbs free energy $G(\mu,H_1,H_2)$ from Eq.~(\ref{TBA-G}), we observe that the susceptibilities $\chi_1$ and $\chi_2$ satisfy the relation 
 (\ref{susceptibility}) in the TLL phase of trions, pairs and single atoms, see Fig.~\ref{sus-H2}.

Using the TBA equations \cite{Tak99,LeeGBY11},  one  can  easily prove that the specific heat can be written in terms of  sound velocities, i.e.\ 
\begin{eqnarray}
c_{V}= \frac{\pi^{2} k_{\rm B}^{2} T}{3} \sum_{r}^{w} \frac{1}{\hbar \pi v_r}. \label{F2-cv}
\end{eqnarray}

\section{Scaling forms on boundaries}
\label{App-II}

We focues on the low temperature behaviour for $w=3$ to derive the analytical results for the thermodynamics, and finally we derive formula of general SU(w) gases of the cricarity about susceptibilities.In the strong interaction limit, we can solve these TBA equations analytically at low temperatures $T\ll \epsilon_3/k_{\rm B}$ by simplifying 
\begin{equation}
\varepsilon _{r}(k) \approx  r\,k^{2}-A_{r},\qquad r=1,\,2,\,3,
\label{TBA-simplified}
\end{equation}
with
\begin{eqnarray}
A_{1} &=&\mu +H_{1}-\frac{2}{|c|}p_{2} -\frac{2}{3|c|}p_{3} +
\frac{1}{4|c|^3}Y_{2,\frac{5}{2}} \nonumber\\
&&+\frac{1}{9|c|^3}Y_{3,\frac{5}{2}}+Te^{-(2H_1-H_2)/T}e^{-J_1/T}I_0(J_1/T) \nonumber \\
A_{2} &=&2\mu +\frac12 {c^{2}}+H_{2}
-\frac{4}{|c|}p_{1}-\frac{1}{|c|}p_{2}- \frac{16}{9|c|}p_{3} \nonumber\\
&&+
\frac{8}{|c|^3}Y_{1,\frac{5}{2}}+
\frac{1}{4|c|^3}Y_{2,\frac{5}{2}} +
\frac{224}{243|c|^3}Y_{3,\frac{5}{2}}\nonumber\\
&&+Te^{-(2H_2-H_1)/T}e^{-J_2/T}I_0(J_2/T),  \nonumber\\
A_{3} &=&3\mu +2c^{2}-\frac{2}{|c|}p_{1} -\frac{8}{3|c|}p_{2}
-\frac{1}{|c|} p_{3} \nonumber\\
&&+ \frac{1}{2|c|^3}Y_{1,\frac{5}{2}}+ \frac{28}{27|c|^3}Y_{2,\frac{5}{2}} +
\frac{1}{16|c|^3}Y_{3,\frac{5}{2}}
\label{TBA-simplified2}
\end{eqnarray}
with  $Y_{r,a}=-\sqrt{\frac{r}{4\pi}}T^a{\rm Li}_a\left(-e^{A_{r}/T}\right)$.
Here the effective spin coupling constant is $J_r=2p_{r}/r|c|$ with $c=mg_{1D}/\hbar ^{2}$. The polylogarithm function is
defined as $\mathrm{Li}_{n}(x)=\sum_{k=1}^\infty \frac{x^k}{k^n}$ and $I_0(x)=\sum_{k=0}^{\infty} \frac{1}{(k!)^2}\left(x/2\right)^{2k}$.
The effective pressures $p_{r}$, with $r=1$, $2$, $3$, of excess fermions, pairs and trions, respectively, can be expressed as 
\begin{eqnarray} 
p_{1} &=&  Y_{1,\frac{3}{2}}\left[1+\frac{4}{|c|^3}
Y_{2,\frac{3}{2}} + \frac{1}{3|c|^3}Y_{3,\frac{3}{2}}
\right], \nonumber\\
p_{2}&=& Y_{2,\frac{3}{2}}\left[1+\frac{4}{|c|^3}Y_{1,\frac{3}{2}}
+\frac{1}{4|c|^3}Y_{2,\frac{3}{2}} + \frac{112}{81|c|^3}
Y_{3,\frac{3}{2}}
\right], \nonumber\\
p_{3} &=& 
Y_{2,\frac{3}{2}}\left[1+\frac{1}{3|c|^3}Y_{1,\frac{3}{2}}
+\frac{112}{81|c|^3}Y_{2,\frac{3}{2}} + \frac{1}{8|c|^3}
Y_{3,\frac{3}{2}}\right].\nonumber
\end{eqnarray}

\subsection{Vacuum -- Pair} 
The above simplified TBA equations (\ref{TBA-simplified}) can be used to derive universal low temper properties of the SU(2) and SU(3) Fermi gases. In the following, we derive the scaling forms of the Wilson ratios in the critical regions for  the two-component Fermi gas. Here we will use the dimensionless units as being explained in the main text. 
The phase boundary   for the phase transition from V  to P phase are
$\tilde{\mu}_c = - \frac{1}{2}$ for $h=H/\varepsilon_2 < 1$.
Near this  critical point, the scaling forms of specific heat, compressibility and susceptibility in dimensionless units are given by
\begin{eqnarray}
  \frac{c_V}{| c | t} &\approx& - \frac{2}{\sqrt{\pi t}} \mathcal{G} \left(
  \frac{2 ( \tilde{\mu} - \tilde{\mu}_c)}{t} \right),\\
  \tilde{\kappa} &\approx &- \frac{2}{\sqrt{\pi t}} \mathcal{F} \left( \frac{2 (
  \tilde{\mu} - \tilde{\mu}_c)}{t} \right),\,\,\,  \\
  \tilde{\chi} & \approx &  0.
\end{eqnarray}
where the functions $\mathcal{G} ( x) = \frac{3}{16} {\rm Li}_{\frac{3}{2}} ( - e^x) -
\frac{1}{4} x {\rm Li}_{\frac{1}{2}} ( - e^x) + \frac{1}{4} x^2 {\rm Li}_{-
\frac{1}{2}} ( - e^x)$ and $\mathcal{F}  ( x) = {\rm Li}_{- \frac{1}{2}} ( -
e^x)$.

\subsection{Vacuum -- Fully Polarized} 

The phase boundary  for the phase transition from V to F phase are
$\tilde{\mu}_c = - \frac{h}{2}$, for $h > 1$;
 the scaling forms in dimensionless units are 
 \begin{eqnarray}
  \frac{c_V}{| c | t}& \approx& - \frac{1}{2 \sqrt{2 \pi t}} \mathcal{H} \left(
  \frac{\tilde{\mu} - \tilde{\mu}_c}{t} \right),\\
  \tilde{\kappa} &\approx &- \frac{1}{2 \sqrt{2 \pi t}} \mathcal{F} \left(
  \frac{\tilde{\mu} - \tilde{\mu}_c}{t} \right),\\
  \tilde{\chi} &\approx &- \frac{1}{8 \sqrt{2 \pi t}} \mathcal{F} \left(
  \frac{\tilde{\mu} - \tilde{\mu}_c}{t} \right),
\end{eqnarray}
where $\mathcal{H} ( x) = \frac{3}{4} {\rm Li}_{\frac{3}{2}} ( - e^x) - x
{\rm Li}_{\frac{1}{2}} ( - e^x) + x^2 {\rm Li}_{- \frac{1}{2}} ( - e^x)$.

\subsection{Pair -- Partially Polarized} 

The critical  fields for the phase transition  from P to F+P  phase are
$\tilde{\mu}_c = - \frac{h}{2} + \frac{4}{3 \pi} ( 1 - h)^{\frac{3}{2}}$, for $h < 1$;
 the scaling forms are
\begin{eqnarray}
  \frac{c_V}{| c | t} & \approx & - \frac{1}{2 \sqrt{2 \pi t}} \mathcal{R} \left(
  \frac{\tilde{\mu} - \tilde{\mu}_c}{t} \right),\\
 \tilde{\kappa} &\approx&  \kappa_{o 1} - \frac{\lambda_1}{2 \sqrt{2 \pi t}}
  \mathcal{F} \left( \frac{\tilde{\mu} - \tilde{\mu}_c}{t} \right),\\
  \tilde{\chi} &\approx & - \frac{\lambda_2}{8 \sqrt{2 \pi t}} \mathcal{F} \left(
  \frac{\tilde{\mu} - \tilde{\mu}_c}{t} \right),
\end{eqnarray}
where $b = ( 1 - h) \left( 1 + \frac{2}{\pi} \sqrt{1 - h} \right), \lambda_1 =
1 + \frac{2 \sqrt{b}}{\pi} - \frac{10 b}{\pi^2}$, $\lambda_2 = 1 - \frac{3
\sqrt{b}}{\pi} + \frac{6 b}{\pi^2}$, $\kappa_{o 1} = \frac{2}{\pi \sqrt{b}}
\lambda_1$ and $\mathcal{R} ( x) = \frac{3}{4} {\rm Li}_{\frac{3}{2}} ( - e^x)
- x {\rm Li}_{\frac{1}{2}} ( - e^x) + x^2 {\rm Li}_{- \frac{1}{2}} ( -
e^x)$.

\subsection{Fully -- Partially Polarized} 

For the phase transition  from F to F+P  phase, we have
$\tilde{\mu}_c = - \frac{1}{2} + \frac{1}{3 \pi} ( h - 1)^{\frac{3}{2}}$, $h > 1$; 
with scaling forms
\begin{eqnarray}
  \frac{c_V}{| c | t} & \approx& - \frac{1}{2 \sqrt{2 \pi t}} \mathcal{S} \left(
  \frac{2 ( \tilde{\mu} - \tilde{\mu}_c)}{t} \right),\\
  \tilde{\kappa} &\approx & \kappa_{o 3} - \frac{\lambda_3}{2 \sqrt{2 \pi t}}
  \mathcal{F} \left( \frac{2 ( \tilde{\mu} - \tilde{\mu}_c)}{t} \right),\\
  \tilde{\chi} &\approx&  \chi_{o 4} - \frac{\lambda_4}{8 \sqrt{2 \pi t}}
  \mathcal{F} \left( \frac{2 ( \tilde{\mu} - \tilde{\mu}_c)}{t} \right).
\end{eqnarray}
where the constants are given by
$a = ( h - 1) \left( 1 + \frac{2}{3 \pi} \sqrt{h - 1} \right)$,
$\lambda_3 = 4 \sqrt{2} \left( 1 + \frac{\sqrt{a}}{\sqrt{2} \pi} - \frac{a}{\pi^2} \right)$,
$\lambda_4 = \frac{8 \sqrt{2} a}{\pi^2}, \,\,\,\kappa_{o 3} = \frac{1}{2\sqrt{2} \pi \sqrt{a}}$,  and
$\chi_{o 4} = \frac{1}{8 \sqrt{2} \pi \sqrt{a}} \left( 1 - \frac{3 \sqrt{a}}{\pi} \right)$.
The dimensionless function is given by
$S(x) = \frac{3}{2 \sqrt{2}} {\rm Li}_{\frac{3}{2}} ( - e^x) - \sqrt{2} x {\rm Li}_{\frac{1}{2}} ( - e^x) +\sqrt{2} x^2 {\rm Li}_{- \frac{1}{2}} ( -e^x)$.
%
The slopes of the Wilson ratio curves at the critical point $\mu=\mu_c$ reveal a unique temperature-dependent feature, namely the slope at the critical point,  
$\left( \frac{\partial R_{\rm w}^{\rm \kappa} }{\partial \mu} \right)_{\mu_c} \equiv  \frac{C_r}{T}$ is given as
\begin{equation}
c_r=\frac{r \pi^2}{3} \frac{ \left( \mathcal{F}' (0) \mathcal{G}(0)-\mathcal{F} (0) \mathcal{G}'(0) \right) }{\mathcal{G}(0)^2},
\end{equation}
 i.e. is a constant for the phase transition from vacuum into an $r$-complex  TLL phase.

\subsection{Susceptibilities for the gaped phase  in the attractive $SU(w)$ gases } 

The total polarization of $SU(w)$ gas can be expressed as
$\tilde{m}=\sum_{r=1}^w\frac{1}{2}\tilde{n}_r r(w-r)$, then we can obtain the susceptibility:
\begin{align}
\tilde{\chi}_r=\frac{\partial \tilde{m}}{\partial \tilde{h}_r}=
\sum_{k,l=1}^w \frac{1}{2}k(w-k) \frac{\partial^2 \tilde{p}^{(l)}}{\partial \tilde{h}_k \partial \tilde{h}_r} \label{chi-r-definition}
\end{align}
in the limit of $t\rightarrow0$ and $|c|>>1$, the leading behavior of the second order derivatives of pressures reads
\begin{align}
\frac{\partial^2 \tilde{p}^{l}}{\partial \tilde{h}_k \partial \tilde{h}_r}
\approx -\frac{\sqrt{r}}{2}\frac{t^{-\frac{1}{2}}}{\sqrt{\pi}}\rm{ Li}_{-\frac{1}{2}}(-e^{\frac{\tilde{A}^{(r)}}{t}})
\delta_{l,k}\delta_{l,r} \label{dp-dhdh-leadingterm}.
\end{align}
Substituting (\ref{dp-dhdh-leadingterm}) into (\ref{chi-r-definition}), we arrive at the explicit form of the
susceptibility $\tilde{\chi}_r$  corresponding  to the field $H_r$ when the other fields are fixed
\begin{align}
\tilde{\chi}_r &\approx -\frac{t^{-\frac{1}{2}}}{4\sqrt{2\pi}}\sqrt{r} (w-r)r
\,\rm{ Li}_{-\frac{1}{2}}(-e^{\frac{\tilde{A}^{(r)}}{t}}) \notag \\
&\approx \frac{t^{-\frac{1}{2}}}{4\sqrt{2\pi}}\sqrt{r} (w-r)r 
\, e^{-\frac{\Delta_r}{t}},  \label{chi-r-result}
\end{align}
where the gap $\Delta_r$ is related to the effective chemical potential through
 $\Delta_r=-\tilde{A}^{(r)}$, which 
can be determined from the TBA equations
\begin{align}
\tilde{A}^{(r)}=r \tilde{\mu}-\sum_{m=1}^w\sum^{\mathrm{min}(r,m)}_{\tiny \begin{array}{c}q=1\\
2q\ne r+m \end{array} }\frac{4\tilde{p}^{m}}{m(r+m-2q)}.
\end{align}
Here the second approximately equal in (\ref{chi-r-result}) holds when $\Delta_r>0$ which implies the 
existence of the gap and the susceptibility presents exponential decay, otherwise $\tilde{\chi}_r \approx 
\frac{\sqrt{r}(w-r)r}{4\sqrt{2}\pi\sqrt{-\Delta_r}}$, which is a positive constant when $t \rightarrow 0$. 
Note that the result in (\ref{chi-r-result}) is different by a factor $\frac{1}{2}$ due to the 
convention  of $H_1=H/2$ in the SU(2) case.
%

\section{Explicit formulas for the SU($2$)  Fermi gas}
\label{App-IV}

In the mixed phase for SU($2$) Fermi gas with one magnetic field $H_1$, we have $n=2n_2+n_1$ and 
$\chi_{1,1} =(\mu_{\rm B} \mathfrak{g})^{2}  \left( {\partial n_{1}}/{\partial \mu_{1}}\right)_{n}$, 
$\chi_{1,2} =2(\mu_{\rm B} \mathfrak{g} )^{2} \left( {\partial n_{2}}/{\partial \mu_{2}}\right)_{n}$, with the stiffnesses in canonical ensemble
$D_1^\chi = \frac{1}{\hbar  \pi} \left( {\partial \mu_{1}}/{\partial n_{1}} \right)_{n }$,
$D_2^\chi = \frac{2}{\hbar \pi} \left( {\partial \mu_{2}}/{\partial n_{2}} \right)_{n}$.
Similarly, 
$\kappa_{ 1} = \left( {\partial n_{1}}/{\partial \mu_{1}} \right)_{H}$ and
$\kappa_{ 2} = 2\left( {\partial n_{2}}/{\partial \mu_{2}} \right)_{H}$ with
$D_2^\kappa = \frac{2 }{\hbar \pi } \left( {\partial \mu_{2}}/{\partial n_{2}} \right)_{H}$, 
$D_1^\kappa = \frac{1}{\hbar   \pi} \left( {\partial \mu_{1}}/{\partial n_{1}} \right)_{H}$ defined in the grand canonical ensemble.

For strong coupling, we find the explicit form of the $\mu_r$ in terms of the densities of pairs $n_2$ and unpaired fermions $n_1$ (in units of $\hbar^2/(2m)$) \cite{LeeGBY11} to be
\begin{eqnarray}
 {\mu}_2 &\approx & \pi^2 \left( \frac{n_2^2}{4} + \frac{2 n_2^3}{3 | c |} +
  \frac{n_2^2 n_1}{| c |} + \frac{4 n_1^3}{3 | c |} + \frac{3 n_1^2
  n_2^2}{c^2} + \frac{5 n_2^4}{4 c^2} \right. \nonumber \\
  && \left.+ \frac{4 n_1 n_2^3}{c^2} + \frac{16
  n_1^3 n_2}{c^2} \right), \\
  {\mu}_1& \approx & \pi^2 \left( n_1^2 + \frac{8 n_1^2 n_2}{| c |} + \frac{2
  n_2^3}{3 | c |} + \frac{48 n_1^2 n_2^2}{c^2} + \frac{4 n_2^3 n_1}{c^2} \right. \nonumber \\
  &&  \left. +\frac{2 n_2^4}{c^2} \right) .
\end{eqnarray}
Here $c=mg_{\rm 1D}/\hbar^2=-2/a_{\rm 1D}$ is the interaction strength.
Therefore, the compressibilities and susceptibilities are given by (in units of $\hbar^2 /(2m)$), e.g.\ 
\begin{eqnarray}
  \kappa_{\mathrm{2}}^{- 1} & \approx & \frac{\pi^2 n_2}{4}  \left( 1 +
  \frac{6 n_1}{|c|} + \frac{4 n_2}{|c|} + \frac{n_2^2}{2 |c|n_1} + \frac{24
  n_1^2}{c^2}  \right.
  \nonumber\\
  &  & \left.+ \frac{24 n_1 n_2}{c^2} + \frac{17 n_2^2}{c^2} - \frac{2 n_2^3}{c^2 n_1} + \frac{n_2^4}{4 c^2 n_1^2} \right),\label{com-ub}
  \\
  \kappa_{\mathrm{1}}^{- 1} & \approx & 2 \pi^2 n_1  \left( 1 + \frac{12
  n_2}{|c|} + \frac{16 n_1^2}{|c|n_2} + \frac{96 n_2^2}{c^2} + \frac{384
  n_1^2}{c^2}  \right.\nonumber \\
 && \left.- \frac{8 n_1 n_2}{c^2}- \frac{96 n_1^3}{c^2 n_2} + \frac{256
  n_1^4}{c^2 n_2} \right)\\
  \chi_{1,1}^{-1}&=&  \pi^2 n_2\left[ 1+\frac{4}{|c|}(n-3n_2)+\frac{3}{c^2}(4n^2\right.\nonumber \\
  &&\left. -24nn_2+30n_2^2)\right],\\
  \chi_{1,2}^{-1}&=&  8\pi^2 n_1 \left[ 1+\frac{4}{|c|}(n-2n_1)+\frac{4}{c^2}(2n^2\right.\nonumber\\
  &&\left. +10n_1^2-12nn_1)   \right].
\end{eqnarray}
In the above calculation for compressibility, the condition  $H = 2 ({\mu}_{\rm 1} -{\mu}_{\rm 2}) + c^2 / 4$ was used, i.e.\ $dH=0$ gives (up to the $ O \left( \frac{1}{c^2} \right)$ order):
\begin{eqnarray}
  \frac{d n_1}{d n_2} = \frac{n_2}{4 n_1} \left( 1 - \frac{8 n_2}{| c |} +
  \frac{6 n_1}{| c |} + \frac{n_2^2}{2 |c|n_1} - \frac{16 n_1^2}{|c|n_2} \right).
  \label{dH}
\end{eqnarray}
Moreover,  the interaction effect enters into the  collective velocities $v_{1}$, $v_{2}$  of the excess fermions and bound pairs.
For strong attraction, they  are given by \cite{GuaBL13}
\begin{eqnarray} \label{equ:sound_velecity_1}
v_{1}&\approx &\frac{\hbar }{2m}2\pi n_{1}   \left(1+ 8 n_{2} /|c| +48 n_{2}^2/c^2 \right),\nonumber \\
v_{2}&\approx &\frac{\hbar }{2m} \pi n_{2}  \left(1+ 2A/|c| +3A^2/c^2 \right),
\end{eqnarray}
with $A=2n_{1}+n_{2}$.

\section{Connection between the Luttinger and  Fermi liquids}

{\bf Wilson Ratios.} Now we further build up a connection between  the TLL and the Fermi liquid. 
By definition (\ref{eq-WR-chi})  and (\ref{eq-WR-kappa}), the two type of Wilson ratios of interacting Fermi liquid  in 3D are given by 
\begin{eqnarray}
R^{\chi}_{\rm W}=\frac{1}{1+F^a_0},\,\,\, R^{\kappa }_{\rm W}=\frac{1}{1+F^s_0}\label{WR-FL}
\end{eqnarray}
which  depends on Landau parameters $F_0^{a,s}$ charactering the interaction \cite{Hew97}. 
This is very similar to our finding for 1D systems, see Eqs. (\ref{WR1}) and (\ref{WR2}). 
However, it's extremely hard to calculate the Landau parameters $F^{s,a}_{0,1}$ in Fermi liquid theory due to  the reason that $N^\ast(0)$ can not be obtained explicitly.
Fermi liquid theory elegantly maps an interacting system into a free fermion system where the interaction is encoded into the density of state and effective mass. 
But the  Fermi liquid theory   is not valid in 1D interacting systems   because  there does not exist a  well-defined
quasiparticle. Here we demonstrate  that exact solution of the TBA equations does show  a  novel existence of Fermi liquid like signature in 1D interacting systems. 
In fact, the additivity rules which we found in previous sections   reflect  a 1D Fermi liquid like nature, also see a discussion in  \cite{Wan98}.

{\bf Feedback interaction.} The excitations near  Fermi points in 1D many-body systems can form a collective motion independent of microscopic details of systems, namely,  there  exists a certain dispersion relation between energy and momentum. We observe that integrable systems provide a deep  understanding of the  intrinsic connection between the TLL and the Fermi liquid.
In fact, the TBA equations of 1D systems  give  the exact dressed energies and determine  the  dispersion relations of each branch.
At low temperatures, only the behavior of the dressed energies near the zero point, or saying, the $k_F$ and $v_F$ at the 1D Fermi points, determines the first order and second orders of thermodynamic quantities. In general,  for the attractive $SU(N)$ Fermi gases, the spin fluctuations are suppressed. Therefore the TBA  equations can be rewritten as 
\begin{align}
\varepsilon^{(r)}(k)=&\epsilon_r^0(k)-\sum_{s=1}^N A_{rs}\ast \varepsilon_{-}^{(s)}(k), \,\,\; 
(r=1,2,\cdots,w),\nonumber  \\
A_{mn}=&\sum_{q=1}^{min(m,n)} a_{m+n-q}
\label{TBA-FL}
\end{align}
where $\epsilon_r^0=\frac{\hbar^2}{2m}r(k^2-\mu_r)=\frac{p^2_0}{2r m}-\frac{\hbar^2 }{2m}r\mu_r $ is the first order coefficient describing excitation energy of a single $r$-complex, here $\mu_r$  the effective chemical potential. Here $p_0=r \hbar k$. The function $a_m(x)$ is defined in Appendix A.  
These equations show  a similar  form of 'feedback interaction' equation in the  Landau Fermi liquid theory \cite{Sch99,Wan98}.
This encourage us to further find a  perturbation idea into our calculations.
What below  gives a  Fermi liquid like description for the low energies  of  the 1D attractive Fermi gases, i.e. mapping an interacting system with multi-subsystems to a multicomponent free system. 
%

 {\bf Phenomenological description.} The symmetry group of the interaction of quasi particles in the 1D ``Fermi liquid"   is reduced from the $SO(3)$ group  to the cyclic group $C_2$.
Therefore,  interaction parameter between the 'quasi' momentum $p$ and $p^\prime$ could be written as $f^{s,a}_{p,p^\prime}=f^{s,a}_0+f^{s,a}_1 sign(p\cdot p\prime)$. Then following the conventional Fermi liquid  theory,  we still can have Landau parameters $F^{s,a}_{0,1}$, which are interacting parameters \cite{Wan98} and describe the main properties of our system. 
 Conventional sound velocity   is associated with oscillations in the density of a fluid, and hydrodynamics gives
that
\begin{eqnarray}
v^2=\frac{\kappa}{\rho}=\frac{\kappa}{m n}.
\end{eqnarray}
Where $\rho=m n$ is the density of the fluid and $\kappa=-L\frac{\partial P}{\partial L}$ is the bulk
modulus.
In FL theory $\kappa=\frac{n^2}{N^\ast(0)}(1+F^s_0)$, where the $N^\ast(0)$ denotes the density of state in momentum space near Fermi surface,  therefore  the velocity of first sound is given by
\begin{eqnarray}
v^2=\frac{n}{m N^\ast(0)} (1+F^s_0)
\end{eqnarray}
By the  1D analog of Fermi liquid theory, we have the relations \cite{Wan98}
$n=\frac{p_F}{\pi}$, $N^\ast(0)=\frac{1}{\pi}\frac{m^\ast }{p_F}$ and $m=m^\ast (1+F^s_1)$.
Then we obtain:
\begin{eqnarray}
v^2=v_F^2 (1+F^s_0) (1+F^s_1). \label{eq:sound_velocity}
\end{eqnarray}
Here, due to the collective motion, the backflow does not exist in 1D, therefore $F^s_1\approx 0$. This can be  seen from the dressed energy equation (\ref{TBA-FL}), where  the effective mass $r\,m$ is almost unchanged up to the order of $O(c^{-2})$ for a  strong coupling regime.  

{\bf Consistency. } Using the exact solutions, we can calculate
the Wilson ratio $R_W^c$ for  different phases of TLLs.
For a single state of a $r$-complex,  the Wilson ratio is known from Eq.~(\ref{WR-FL}).
Then  the parameter $F^s_{0,1}$ could be determined by a comparison with the Bethe ansatz result in strong coupling limit (i.e. $|c|\to \infty$) via 
\begin{eqnarray}
R_{\rm W}^{\rm c}=\frac{1}{(1+F^s_0)}=
\begin{cases}
 1; &   \quad  \text{for\,  free \, fermions} \\
 4; &   \quad   \text{for\, pairs}\\
 9; &   \quad \text{for \, trions }
\end{cases}. \label{eq:landau_parameters}
\end{eqnarray}
From the equation (\ref{eq:sound_velocity}) and (\ref{eq:landau_parameters}) we could calculate
 the relation between Fermi speed and sound velocity. For example, the velocities of   trions, pairs and single fermionic atoms,  are given by 
\begin{eqnarray}
v^3=\frac{1}{3}v_F^{(3)}; \,\,v^2=\frac{1}{2}v_F^{(2)}; \, \,v^1=v_F^{(1)}, \label{eq:relation_of_vs_vf}
\end{eqnarray}
respectively. This builds up an intrinsic connection between the Luttinger liquid and Fermi liquid. 

Furthermore, from the free fermion nature of the Fermi liquid,  we can express the
specific heat in terms of these   sound velocities
\begin{eqnarray}
 \frac{c_{V }}{T}=\frac{\pi}{3} \left(
   \frac{1}{v_{F}^{ 1 }} + \frac{2}{v_{F}^{ 2 }} + \frac{3}{v_{F}^{ 3 }}
   \right)=\frac{\pi}{3} \left(
   \frac{1}{v^{ 1 }} + \frac{1}{v^{ 2 }} + \frac{1}{v^{ 3 }}
   \right), \label{eq:sumup_specific_heat_sound}
\end{eqnarray}
where we have dropped  the unit $k_{B}^{2}\hbar^{-1}$, here the interaction effect is encoded in the velocities $v^{1,2,3}$.  This equation could be obtained by consiering the leading term of the TBA equations at low temperature region. 
 This  shows a consistency  of our derivation by the 1D Fermi liquid like description. 
The derivation above could be  extended to the SU(w) attractive Fermi gases  in a straightforward way.
All the derivation above could be directly extended to the SU(w) attractive Fermi gases:
\begin{eqnarray}
\frac{c_{V }}{T} = \frac{\pi}{3} \left(
   \frac{1}{v^{ 1 }} + \frac{1}{v^{ 2 }} + \cdots +\frac{1}{v^{ w }}
   \right).  \label{eq:sumup_su(w)_specific_heat_sound}
\end{eqnarray}
We can also prove this additivity rule using the TBA equations, see Appendix B.



\end{document}